\documentclass[12pt,a4paper]{article}
\usepackage[colorlinks=true,linkcolor=blue,urlcolor=blue,filecolor=black,citecolor=red,pdfstartview=FitV,pdftitle={},pdfsubject={},
pdfkeywords={},pdfpagemode=None,bookmarksopen=true]{hyperref}
\usepackage{graphicx}
\usepackage{epstopdf}%
\usepackage{amsmath}
\usepackage{amsfonts}
\usepackage{amssymb}
\usepackage{color}%
\usepackage{dcolumn}
\usepackage{slashed}
\usepackage{amssymb,ulem}
\usepackage{float}
\usepackage{amsthm,amsmath,amssymb}
\usepackage{mathrsfs}
\usepackage{subfigure}
\usepackage{wrapfig}
\usepackage{indentfirst}

\renewcommand{\thefootnote}{\fnsymbol{footnote}}

\begin{document}

\vspace{12mm}

\begin{center}
{{{\Large {\bf Analytical approximations to charged black hole solutions in Einstein-Maxwell-Weyl gravity }}}}\\[10mm]

Sheng-Yuan Li$^{a}$\footnote{lishengyuan314159@hotmail.com;},
Ming Zhang$^{b}$\footnote{corresponding author. mingzhang0807@126.com;}, De-Cheng Zou$^{a,c}$\footnote{corresponding author. dczou@yzu.edu.cn;} and Meng-Yun Lai$^{c}$\footnote{mengyunlai@jxnu.edu.cn;}\\[8mm]
{${}^a$College of Physical Science and Technology, Yangzhou University, \\ Yangzhou 225009, China\\[0pt]}
{${}^b$Faculty of Science, Xi’an Aeronautical University, Xi’an 710077, China\\[0pt]}
{${}^c$College of Physics and Communication Electronics, Jiangxi Normal University, \\ Nanchang 330022, China\\[0pt] }

\end{center}
\vspace{2mm}

\begin{abstract}
The Homotopy Analysis Method (HAM) is a useful method to derive analytical approximate solutions of black holes in modified gravity theories. In this paper, we study the Einstein-Weyl gravity coupled with Maxwell field, and obtain analytical approximation solutions for charged black holes by using the HAM. It is found that the analytical approximate solutions are sufficiently accurate in the entire spacetime outside the black hole's event horizon, and also consistent with numerical ones for charged black holes in the Einstein-Maxwell-Weyl gravity. 

\end{abstract}

\vspace{5mm}

\vspace{1.5cm}

\hspace{11.5cm}
\newpage
\renewcommand{\thefootnote}{\arabic{footnote}}
\setcounter{footnote}{0}


\section{Introduction}

It is widely recognized that the theory of general relativity (GR) does not qualify as a renormalizable quantum field theory within the framework of effective field theory. Therefore, to achieve the ultimate goal of unifying GR with quantum theory, it is imperative to explore alternative theories that go beyond GR. A possible attempt to solve the problem of the non-renormalizability of GR is to include higher-order corrections that become
important at higher energy. Some years ago, Stelle  \cite{Stelle:1976gc} proposed to add all possible
quadratic curvature invariants to the usual Einstein-Hilbert action, and then obtained a theory of quantum gravity
free of ultraviolet divergences but as early recognized \cite{Pais:1950za}, albeit at the price of introducing ghost-like modes. In four-dimensional spacetime, the most comprehensive theory that includes second-order derivative terms in the curvature can be expressed as follows\cite{Holdom:2016nek,Lu:2015cqa}
\begin{equation}
\mathcal{I} =\int d^4 x\sqrt{-g}\left( \gamma R-\alpha C_{\mu\rho\nu\sigma} C^{\mu\rho\nu\sigma} + \beta R^2 \right),\label{Action1}
\end{equation}
where the parameters $\alpha$, $\beta$, and $\gamma$ are constants, and $C_{\mu\rho\nu\sigma} $ is the Weyl tensor. In addition, black holes are fundamental objects in theories of gravity and serve as powerful tools for studying the intricate global aspects of the theory. Subsequently, L{\"u} \textit{et al}.\cite{Lu:2015cqa,Lu:2017kzi} derived numerical solutions of non-Schwarzschild black holes (NSBH) in the Einstein-Weyl gravity [Eq.\eqref{Action1}] by considering the disappearance of the Ricci scalar for any static spherically symmetric black hole solution ($R=0$). Actually, the no-go theorem also implies that the Ricci scalar $R$ must be zero for a black hole in the case of pure gravity or with a traceless matter stress tensor. In Refs.~\cite{Lu:2015cqa,Lu:2017kzi}, they further investigated some thermodynamic properties of NSBH and discovered several remarkable features: 1) the NSBH can have positive and negative masses; 2) as the coupling constant $\alpha$ approaches its extreme value, the black hole tends towards a massless state while maintaining a nonzero radius. Recently, Held \textit{et al}.\cite{Held:2022abx} discussed the linear stability of these two branches of black hole solutions. In addition, charged black holes in the Einstein-Weyl gravity coupled with (nonlinear-) Maxwell field were constructed in Refs. \cite{Lin:2016jjl,Lin:2016kip} and our previous works \cite{Zou:2019onm,Wu:2019uvq}, where two sets of numerical solutions were obtained: the charged generalization of Schwarzschild solution and charged generalization of non-Schwarzschild solution. The analysis of quasinormal modes of the non-Schwarzschild and charged black holes was performed in the Einstein-Weyl gravity \cite{Cai:2015fia,Zinhailo:2018ska,Zou:2020msk}, where the linear relation between quasinormal modes frequencies and the parameter was recovered. Later, the black holes with massive scalar hair were obtained in Ref.~\cite{Sultana:2020pcc}, where it discussed the effects of the scalar field on the black hole structure. Recently, some novel solutions of black holes were also studied in the Einstein-Weyl gravity \cite{Huang:2022urr,Podolsky:2018pfe}, including the phase diagram of Einstein-Weyl gravity \cite{Silveravalle:2022wij}.

However, the numerical solutions for non-Schwarzschild black holes have limitations in providing a clear understanding of the dependence of the metric on physical parameters, as they are obtained at fixed parameter values and displayed as curves in figures rather than explicit expressions. This makes it difficult for researchers to use these solutions in their work without recalculating them. Fortunately, there are general methods available for parametrizing black hole space-times, such as the Continued Fractions Method (CFM)~\cite{Rezzolla:2014mua,Kokkotas:2017zwt} and the Homotopy Analysis Method (HAM)~\cite{Liao:1992mua,Liao:2003mua,Liao:2012mua}. In this paper, we focus on the HAM. It is considered as a very useful method for obtaining analytical approximate solutions for various nonlinear differential equations, including those arising in different areas of science and engineering. Despite its widespread use in other fields, the HAM has been limited in the fields of general relativity and gravitation. Recently, we constructed analytical approximation solutions of scalarized AdS black holes in Einstein-scalar-Gauss–Bonnet gravity by using of HAM \cite{Zou:2023inv}. Moreover, this HAM has been adopted to derive the analytical approximation solutions of non-Schwarzschild black holes in the Einstein-Weyl gravity~\cite{Sultana:2019lhf}, analytic approximate solutions of hairy black holes in Einstein–Weyl-scalar gravity~\cite{Sultana:2021cvq} as well as for the Regge-Wheeler equations under metric perturbations on the Schwarzschild spacetime~\cite{Cho:2020tzx}. 
The above works inspire us to continue exploring the application of homotopy analysis in modified gravity theory. In this work, we wish to apply the HAM to obtain analytical approximation solutions of charged black holes in the Einstein-Maxwell-Weyl gravity.

The plan of the paper is as follows. In Sec.~\ref{sec2} we review the Einstein-Weyl gravity coupled with Maxwell field and present the numerical solutions for charged black hole in the Einstein-Maxwell-Weyl gravity. Sec.~\ref{sec3} is devoted to deriving the analytical approximation solutions by using the HAM method, where two solutions are accurate in the whole space outside the event horizon. The paper ends with a discussion of the results obtained in  Sec.~\ref{sec4}.

\section{The Einstein-Maxwell-Weyl gravity}  \label{sec2}

The action of Einstein-Weyl gravity in the presence of Maxwell field is given by \cite{Lin:2016jjl}
\begin{equation}
\mathcal{I} =\int d^4 x\sqrt{-g}\left( \gamma R-\alpha C_{\mu\rho\nu\sigma} C^{\mu\rho\nu\sigma} + \beta R^2 -\kappa F_{\mu\nu} F^{\mu\nu} \right),\label{Action2}
\end{equation}
where the parameters $\gamma$, $\alpha$, $\beta$, and $\kappa$ are coupling constants, $F_{\mu\nu} =\nabla_{\mu}A_{\nu}-\nabla_{\nu}A_{\mu}$ is the electromagnetic tensor and $C_{\mu\rho\nu\sigma}$ is the Weyl tensor. Since resulting tensors in the equations of motion that comes from the Weyl and Maxwell energy momentum tensors are traceless, a charged black hole solution in this theory should not need of the contribution from $\beta R^2$ term~\cite{Lin:2016jjl}. Taking $\beta=0$ and $\gamma=1$, the corresponding field equations are obtained as
\begin{eqnarray}
&&R_{\mu\nu} - \frac{1}{2} g_{\mu\nu} R -4\alpha B_{\mu\nu}-2\kappa T_{\mu\nu}=0, \label{fieldeq} \\
&&\nabla_{\mu} F^{\mu\nu}=0, \label{Maxwelleq}
\end{eqnarray}
where the trace-free Bach tensor $B_{\mu\nu}$ and energy-momentum tensor of electromagnetic field $T_{\mu\nu}$ are defined as
\begin{eqnarray}
B_{\mu\nu}=\left( \nabla^{\rho}\nabla^{\sigma} + \frac{1}{2} R^{\rho\sigma} \right) C_{\mu\rho\nu\sigma},\label{Bach tensor} \nonumber\\
T_{\mu\nu}=F_{\alpha\mu}{F^{\alpha}}_{\nu}-\frac{1}{4} g_{\mu\nu} F_{\alpha\beta}F^{\alpha\beta}.
\label{mtensor}
\end{eqnarray}

Considering the static and spherical symmetry metric ansatz
\begin{eqnarray}
ds^2=-h(r)dt^2+\frac{1}{f(r)} dr^2+r^2 ({d\theta}^2+\sin^2 \theta {d\phi}^2),
\label{metric}
\end{eqnarray}
and substituting the metric ansatz into the field equations (\ref{fieldeq})(\ref{Maxwelleq}), we obtain three independent equations
\begin{eqnarray}\label{neom}
h''(r)=&&\frac{4h(r)^2\left(1- r f'(r)- f(r)\right)-rh(r)\left(r f'(r)+4 f(r)\right)h'(r)+r^2f(r)h'(r)^2}{2r^2 f(r)h(r)} \nonumber\\
f''(r)=&&\frac{h(r)-f(r) h(r)-r f(r) h'(r)}{\alpha  f(r) \left[2 h(r)-r h'(r)\right]}\nonumber\\
&&-\frac{h(r) \left[3 r^2 f'(r)^2+12 r f(r) f'(r)-4 r f'(r)+12 f(r)^2-8 f(r)\right]}{2 r^2 f(r) \left[r h'(r)-2 h(r)\right]}\nonumber\\
&&+\frac{\kappa  Q_0^2 h(r)}{\alpha  r^2 f(r) \left[r h'(r)-2 h(r)\right]}\nonumber\\
&&+\frac{f(r) \left[r^2 h'(r)^2-r h(r) h'(r)-2 h(r)^2\right]-r h(r) f'(r) \left[r h'(r)+4 h(r)\right]}{2 r^2 h(r)^2}\nonumber\\
A_t'(r)=&&-\frac{Q_0}{r^2}\sqrt{\frac{h(r)}{f(r)}},
\end{eqnarray}
where the prime ($'$) denotes differentiation with respect to $r$, and $Q_0$ denotes electric charge. If $Q_0\rightarrow0$, the equations of motion \eqref{neom} reduce to
those found in Refs. \cite{Lu:2015cqa,Lu:2017kzi,Kokkotas:2017zwt}, and recover the same solutions: the Schwarzschild black holes (SBH) and the non-Schwarzschild black holes (NSBH). 

Now we derive the numerical solutions of charged black holes. Here we suppose that the spacetime has only one horizon to make easier the expansion of $f(r)$, $h(r)$ and $A_t(r)$ around the event horizon $r_0$
\begin{eqnarray}
&&h \left(r\right)=c[\left(r-r_0\right)+h_{2}\left(r-r_0\right)^2 +h_{3}\left(r-r_0\right)^3+...],\label{hfrexpand1}\\
&&f \left(r\right)=f_{1}\left(r-r_0\right)+f_{2}\left(r-r_0\right)^2 +f_{3}\left(r-r_0\right)^3+...,\label{hfrexpand2}\\
&&A_t \left(r\right)=A_{t0}+A_{t1}\left(r-r_0\right)+A_{t2}\left(r-r_0\right)^2 +...,\label{hfrexpand3}
\end{eqnarray}
where $h_i$, $f_i$ and $A_{ti}$ are constant coefficients of the expansions, and $c$ is the arbitrary scaling factor, which we choose such that $t$ is the time coordinate of a remote observer, i.e.
\begin{eqnarray}
\lim_{r\rightarrow\infty}h(r)=1.
\end{eqnarray}
Moreover, the expansion parameter $A_{t0}$ will be determined by the behavior of vector potential $A_t(r)$ at infinity.

Substituting the above expansions \eqref{hfrexpand1}\eqref{hfrexpand2}\eqref{hfrexpand3} into field equations (\ref{neom}), arbitrary coefficients $h_j$, $f_j$ and $A_{ti}$ with $j\ge2$ can be expressed in terms of $f_1$, for example, $h_2$, $f_2$, $A_{t2}$ and $A_{t1}$ are expressed as
\begin{eqnarray}
&&h_{2}=\frac{1-2f_1 r_0}{r_0^2}-\frac{r_0^2-f_1 r_0^3-\kappa Q_0^2}{8\alpha f_1 r_0^3}, \nonumber\\
&&f_{2}=\frac{1-2f_1 r_0}{r_0^2}-\frac{3\left(r_0^2-f_1 r_0^3-\kappa Q_0^2\right)}{8\alpha f_1 r_0^3},\nonumber\\
&&A_{t1}=-\frac{Q_0}{r_0^2},\quad A_{t2}=\frac{Q_0}{r_0^3}+\frac{Q_0\left(r_0^3f_1+\kappa Q_0^2-r_0^2\right)}{8\alpha r_0^5 f_1^2} .\label{f2}
\end{eqnarray}

On the other hand, at the radial infinity $(r\rightarrow\infty)$, the metric functions
and vector potential can be expanded in power series, this time in terms of $1/r$.
Demanding that the metric components reduces to those of the asymptotically flat
Minkowski spacetime
\begin{eqnarray}
&&h(r)=1-\frac{2M}{r}+...,\quad f(r)=1-\frac{2M}{r}+...,\nonumber\\
&&A_t(r)=\Phi_{\infty}+\frac{Q_0}{r}+...    .\label{infty}
\end{eqnarray}

\begin{figure}[H]
\centering
\includegraphics[scale=0.5]{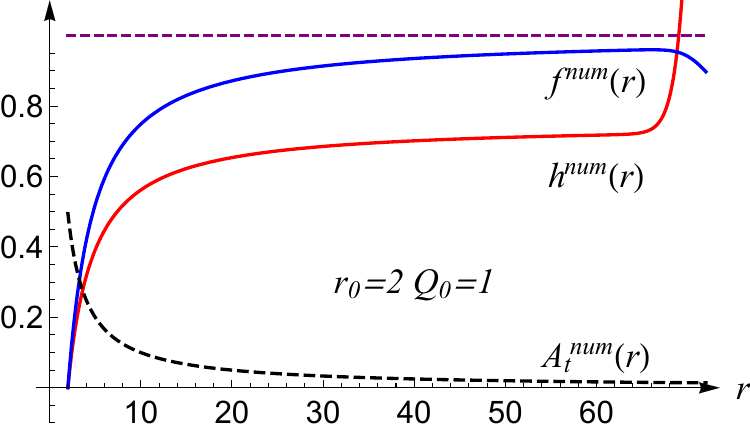}
\quad
\includegraphics[scale=0.5]{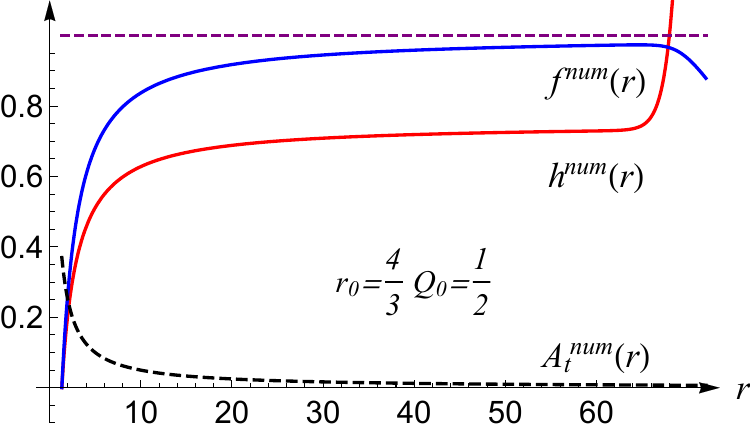}
\caption{Numerical solutions of charged black holes for $h\left(r\right)$, $f \left(r\right)$, and $A_t\left(r\right)$ with different values of $r_0$ and $Q_0$. Here $h\left(r\right)$ has been rescaled to approach $3/4$ to avoid overlap.}
\label{numfig}
\end{figure}

Taking $\alpha = \frac{1}{2}$ and $\kappa = 1$, we assume the initial values of the parameters $Q_0$ and $f_1$ at a radius $r_i=r_0+\frac{1}{1000}$ just outside the horizon $r_0$, and then use numerical routines in $Mathematica\circledR$ to integrate the
equations out to large radius, so that these interpolation
functions of metric functions $h(r)$ and $f(r)$ and vector potential $A_t(r)$ satisfy the boundary condition \eqref{infty}. To ensure that $h(r)$ is asymptotically flat $ {\lim_{r\rightarrow+\infty}} h(r) = 1$, the value of scaling parameter $c$ equals to $0.3367$ for $Q_0=1$ and $r_0 = 2$, and $0.5521$ for $Q_0=1/2$ and $r_0=4/3$, respectively. 
In addition, the vector potential should satisfy the boundary
\begin{eqnarray}
\lim_{r\rightarrow\infty}A_{t}(r)=0.
\end{eqnarray}
For simplicity, we choose $\Phi_{\infty}= 0$ . Then, we can obtain the values of $A_{t0}=0.4950$ and $0.3670$ for charged black holes with $Q_0=1$ and $Q_0=1/2$, respectively. The corresponding numerical solutions are plotted in Fig.~\ref{numfig}.

\section{Analytically approximate solutions}  \label{sec3}

In general, it is a difficult task to find exact solutions of nonlinear differential equations. In Refs.~\cite{Liao:2003mua,Liao:2012mua,Liao2004}, the Homotopy Analysis Method(HAM) was developed to obtain analytical approximate solutions to nonlinear differential equations. In this section, we will derive the analytical approximate solutions of metric functions and the electric potential function by using the HAM.

Consider $n$-nonlinear differential equations system, where $y_i(t)$ is the solution of the nonlinear operator $N_i$ as a function of $t$
\begin{eqnarray}
  N_i[y_i(t)]=0, \qquad\quad i=1,2,...,n, \label{Nequation}
\end{eqnarray}
with unknown function $y_i(t)$ and a variable $t$. Then, the zero-order deformation equation can be written as 
\begin{eqnarray}
(1-q)L[\phi_i(t;q)-y_{i0}(t)]=q h_i H_i(t)N_i[\phi_i(t;q)].\label{hm}
\end{eqnarray}
The HAM constructs a topological homotopy for linear auxiliary operator $L$ and nonlinear operator $N_i$. Introducing an embedding parameter $q\in[0,1]$, when $q$ continuously changes from $0$ to $1$, the solution of the entire equation will also continuously change from the solution $y_{i0}(t)$ of our selected linear auxiliary operator $L$ to the solution $y_i(t)$ of the nonlinear equation $N_i$. In order to control the convergence of the solution, an auxiliary function $H_i(t)$ and a convergence control parameter $h_i$ are also introduced into the homotopy equation (\ref{hm}). By selecting appropriate auxiliary function $H_i(t)$ and convergence control parameter $h_i$, the solution can converge more rapidly.

To decompose the nonlinear problem into a series of linear subproblems, now make Taylor expansions of $\phi_i(t;q)$ with respect to $q$ around $q = 0$
\begin{eqnarray}
\phi_i(t;q)=y_{i0}(t)+\sum_{m=1}^{\infty}y_{im}(t)q^m.\label{phiexpand}
\end{eqnarray}
Where the coefficient $y_{im}(t)$ of the m-th order of $q$ is
\begin{eqnarray}
y_{im}(t)=\frac{1}{m!}\frac{\partial^m \phi_i(t;q)}{\partial q^m}.\label{coeffyim}
\end{eqnarray}
When $q = 1$, it is the expansion of the solution of the nonlinear equation (\ref{Nequation})
\begin{eqnarray}
  y_i(t)=\phi_i(t;1)=y_{i0}(t)+\sum_{m=1}^{\infty}y_{im}(t).
\end{eqnarray}

By solving $y_{im}(t)$, the expansions of the solutions of the nonlinear equations can be found. To achieve it, the operation for the zero order deformation equation (\ref{hm}) will be as follows: First, substitute the expansion (\ref{phiexpand}) into (\ref{hm}). Second, take the m-th derivative of $q$ on (\ref{hm}) both sides. Third, after calculating the derivatives, set $q=0$. The so-called higher order deformation equation ($m$th-order deformation equation) is obtained, which is summarized as
\begin{eqnarray}
  L[y_{im}(t)-\chi_my_{im-1}(t)]=h_i H_i(t) R_{im}(y_{im-1}).
  \label{m-orderhm}
\end{eqnarray}
Where the term $R_{im}$ on the right-hand side with respect to the nonlinear operator is
\begin{eqnarray}
\label{Rim}
  R_{im}(y_{im-1})=\frac{1}{(m-1)!}\frac{\partial^{m-1} N_i[\sum_{m=0}^{\infty}y_{im}(t)q^m]}{\partial q^{m-1}}\mid_{q=0},
\end{eqnarray}
The highest term on the right-hand side of equation (\ref{Rim}) can only reach up to the $y_{im-1}$ term. It is found that $y_{im}(t)$ is related to $y_{im-1}$, and it is possible to find $y_{im}(t)$ of arbitrary order based on this relation. The value of constant $\chi_m$ is
\begin{eqnarray}
  \chi_m=\left\{
  \begin{aligned}
    0 & : m\leq 1, \\
    1 & : m>1 .
  \end{aligned}
  \right.
\end{eqnarray}
In the calculation, we take a finite order to ensure that the error is small enough, a finite M-order approximation
\begin{eqnarray}
\label{m-y-expand}
  y_i^M(t)=y_{i0}(t)+\sum_{m=1}^{M}y_{im}(t),
\end{eqnarray}
the $y_i^M(t)$ are the M-th order approximate solution of the original equation (\ref{Nequation}).

The selection of linear auxiliary operator $L$, initial guess $y_{i0}(t)$, auxiliary function $H_i(t)$, and convergence control parameter $h_i$ has great freedom, making homotopy analysis method highly adaptable to different nonlinear problems. However, precisely because there is a great deal of freedom in selecting these quantities, how to choose these quantities more appropriately still requires a theoretical basis, and related work can be found in \cite{Robert2009}\cite{Yinshan2010}.

Next, we will use the HAM to obtain analytically approximation solutions of charged black holes in the Einstein-Maxwell-Weyl gravity. In order to do it, we choose a coordinate transformation $z=1-\frac{r_0}{r}$ , such that the region of $r\rightarrow\infty$ becomes a finite value $z=1$. Then, the field equations under this coordinate transformation become
\begin{eqnarray}
&&h''(z)-\frac{h'(z)^2}{2 h(z)}+\frac{f'(z) h'(z)}{2 f(z)}+\frac{2 h(z) \left[-(z-1) f'(z)+f(z)-1\right]}{(z-1)^2 f(z)}=0,\label{Neq1}\\
&&f(z) f''(z) \left[2 \alpha  (z-1)^5 h(z)^2 h'(z)+4 \alpha  (z-1)^4 h(z)^3\right]\nonumber\\
&&+f'(z) \Big\{f(z) \left[\alpha  (z-1)^5 h(z) h'(z)^2+2 \alpha  (z-1)^4 h(z)^2 h'(z)+12 \alpha  (z-1)^3 h(z)^3\right]\nonumber\\
&&-4 \alpha  (z-1)^3 h(z)^3\Big\}-3 \alpha  (z-1)^4 h(z)^3 f'(z)^2+f(z) \left[-2 {r_0}^2 (z-1) h(z)^2 h'(z)\nonumber\right.\\
&&\left.+2 {r_0}^2 h(z)^3+8 \alpha  (z-1)^2 h(z)^3\right]+f(z)^2 \left[-\alpha  (z-1)^5 h'(z)^3-3 \alpha  (z-1)^4 h(z) h'(z)^2\nonumber\right.\\
&&\left.-8 \alpha  (z-1)^2 h(z)^3\right]+2 \kappa  {Q_0}^2 (z-1)^2 h(z)^3-2 {r_0}^2 h(z)^3 = 0,\label{Neq2}\\
&&\frac{{Q_0} \sqrt{\frac{h(z)}{f(z)}}}{{r_0}}+{A_t}'(z) = 0,\label{Neq3}
\end{eqnarray}
where the prime ($'$) denotes the differentiation of the function with respect to $z$, and $r_0$ the event horizon of black hole. Notice that Maxwell equation \eqref{Neq3} is a linear one, and \eqref{Neq1} \eqref{Neq2} are second-order derivative equations with respect to $h(z)$ and $f(z)$. Therefore, we derive $h(z)$ and $f(z)$ by applying the HAM to the two nonlinear equations \eqref{Neq1} and \eqref{Neq2}, after then solve equation \eqref{Neq3} for $A_t(z)$.

In the homotopy equation, the auxiliary function can be coded into the initial guess solution, i.e., the $H_i(z)$ on the right side of the zero-order deformation Eq.~\eqref{hm} can be moved to the left side \cite{Robert2009}, so without loss of generality we take the auxiliary function $H_i(z)=1, i=1,2$. The initial guess is
\begin{eqnarray}
&&h_{0}(z)=f_{0}(z)=z\left(1-a\left(1-z\right)\right),A_{t0}(z)=\frac{Q_0}{r_0}(1-z).\label{initials}
\end{eqnarray}
The corresponding linear auxiliary linear operators (whose construction method is shown in \cite{Robert2009}\cite{Yinshan2010})
\begin{eqnarray}
L_i\left[\phi_i(z;q)\right]=\frac{z^2}{2}\frac{\partial^2\phi_i(z;q)}{\partial z^2}-z\frac{\partial\phi_i(z;q)}{\partial z^2}+\phi_i(z;q), \quad i=1,2. \label{auxiliarys}
\end{eqnarray}

\begin{eqnarray}
L_3\left[\phi_3(z;q)\right]=\frac{\partial\phi_3(z;q)}{\partial z}. \label{V-auxiliarys}
\end{eqnarray}

The following boundary conditions are used in the process of solving the nonlinear equations 
\begin{eqnarray}
h(0)=0=f(0), \quad h(1)=1=f(1),\quad A_t(1)=0. \label{bcs}
\end{eqnarray}
The chosen of the initial guess solutions (\ref{initials}) is required to satisfy the boundary conditions (\ref{bcs}).

The nonlinear equations (\ref{Nequation}) $N_i$ is provided by Eqs.~(\ref{Neq1}) and (\ref{Neq2}), and $M$-order analytical approximate solution (\ref{m-y-expand}) is obtained by solving $y_{im}(z)$ from the higher order deformation equation (\ref{m-orderhm}). Here, we do the second-order approximation $M=2$, and the result is related to the parameter $a$ of the initial guessed solution and the convergence control parameters $h_1$, $h_2$ and $h_3$. We assume the convergence control parameters such that $h_3=h_2=h_1$ for simplicity, and then derive the analytical approximate solutions $h(z,a,h_1)$, $f(z,a,h_1)$ and $A_t(z,a,h_1)$, which are related to the undetermined parameters $a$ and $h_1$.

In order to select the optimal values of $a$ and $h_1$, we substitute Eq.\eqref{m-y-expand} into field equations \eqref{Neq1} \eqref{Neq2} \eqref{Neq3}, and obtain
\begin{eqnarray}
\label{delta eq1}
&&\Delta eq_1=|N_1[h(z,a,h_1),f(z,a,h_1)]|,\\
\label{delta eq2}
&&\Delta eq_2=|N_2[h(z,a,h_1),f(z,a,h_1)]|,\\
\label{delta eq3}
&&\Delta eq_3=|N_3[h(z,a,h_1),f(z,a,h_1),A_{t}(z,a,h_1)]|,
\end{eqnarray}
which represented as the deviations between the analytical approximate solutions and the exact solutions.
If the above three equations are as close to zero as possible in $z\in[0,1]$, then that means the approximate solutions are as close as possible to the analytical solutions. 

Now we can evaluate the averaged square residual error \cite{Liao:2012mua,Huang:2010mua} to represent the total deviation between the approximate solutions and the exact solutions
\begin{eqnarray}
E(a,h_1)=&&\frac{1}{S+1}\sum_{k=0}^{S}\Big\{(N_1[h(z_k,a,h_1),f(z_k,a,h_1)])^2+(N_2[h(z_k,a,h_1),f(z_k,a,h_1)])^2\nonumber\\
&&+(N_3[h(z_k,a,h_1),f(z_k,a,h_1),A_{t}(z_k,a,h_1)])^2\Big\},
\label{totalerror}
\end{eqnarray}
with
\begin{eqnarray}
z_k=k\Delta z=k\frac{1}{S}, k=0,1,2,...,S.
\label{zk}
\end{eqnarray}
Then, the optimal values of $a^*$ in the initial guess and convergence control parameter $h_1$ can be obtained from the averaged square residual error, which makes the solution converge more effectively. $a^*$ and $h_1^*$ are points that minimized the averaged square residual error, and the optimal value of $a^*$ and $h_1^*$ is mathematically represented as
\begin{eqnarray}
\{a^*,h_1^*\}=min\{E(a,h_1)\}.
\label{optimal a}
\end{eqnarray}

\begin{figure}[H]
\centering
\subfigure[$r_0 = 2, Q_0 = 1$]{
\label{logtotalerrorfig_a} 
\includegraphics[scale=0.5]{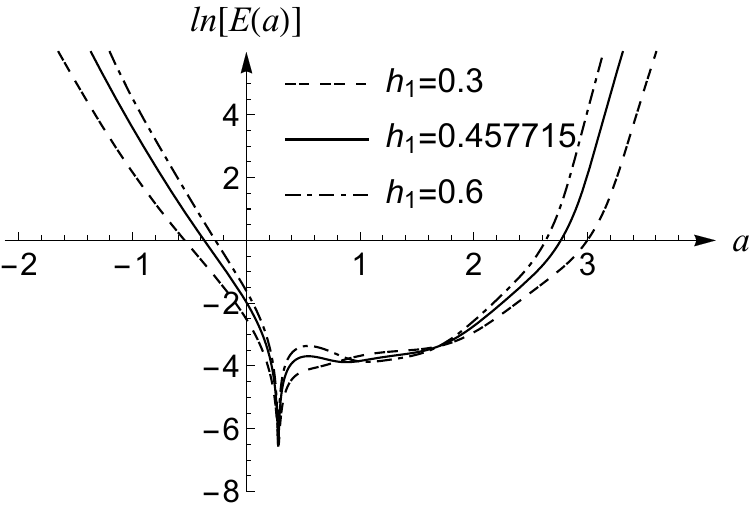}}
\quad
\subfigure[$r_0 =4/3, Q_0 =1/2$]{
\label{logtotalerrorfig_b} 
\includegraphics[scale=0.5]{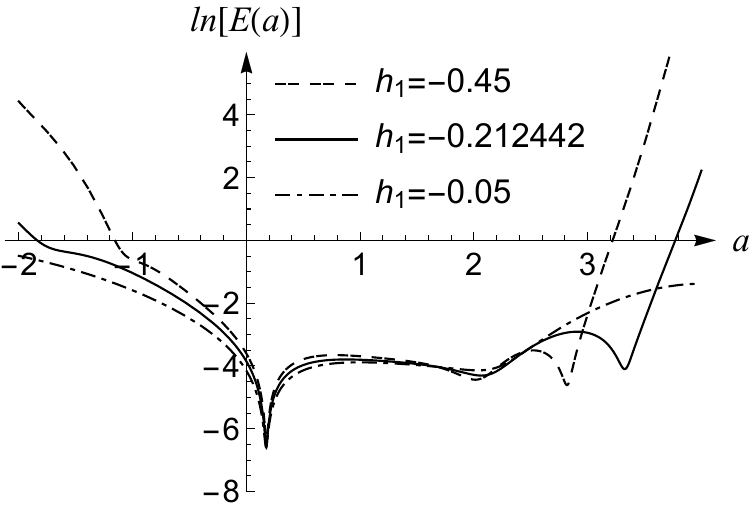}}
\caption{Choose $h_3=h_2=h_1$, and plot the logarithm of the square residual $ln\left[E(a)\right]$ as a function of $a$. (a) When $r_0=2$, $Q_0=1$, the curve of the logarithm of the square residuals, (b)corresponds to $r_0=4/3$, $Q_0=1/2$.}
\label{logtotalerrorfig}
\end{figure}

We choose $S=40$ and use the averaged square residual error (\ref{totalerror}) as functions with the undetermined parameters $a$ and $h_1$. Such an operation may be relegated to symbolic programming with a dedicated ``Minimize'' command in $Mathematica\circledR$. We obtain the optimal convergence-control parameters
$a^*=0.280431$ and $h_1^*=0.457715$, and $a^*=0.174644$ and $h_1^*=-0.212442$ for charged black holes with $Q_0=1$ and $Q_0=1/2$, respectively. Taking different values of $h_1$, the relationships between $E(a)$ and $a$ for $Q_0=1$ and $Q_0=1/2$ are plotted in Fig.~\ref{logtotalerrorfig}.
We find that the minimum points of averaged square residual error $E(a,h_1)$ appear to be weakly sensitive to the parameter $h_1$.

Moreover, considering $Q_0$, $r_0$ and corresponding values of $a^*$ and $h_1^*$, the analytical approximate solutions of charged black holes can be obtained for the different configurations of $Q_0$ and $r_0$. We present an example of $h(r)$, $f(r)$, and $A_t(r)$ for the configuration $Q_0=1$, $r_0=2$, $a^*=0.280431$ and $h_1^*=0.457715$, after reverting back to the radial coordinate $r$
\begin{eqnarray}
\label{hr}
h(r)=
&&\frac{1}{r^{16}}\{-0.0518821+0.820982 r-5.61686 r^2+21.6450 r^3-51.1353 r^4+75.1505 r^5 \nonumber\\
&& -65.8818 r^6+30.0274 r^7-3.91727 r^8-1.57937 r^9+0.435328 r^{10}-0.00210941 r^{11}\nonumber\\
&&+0.00976690 r^{12}+0.00879167 r^{13}+1.10581 r^{14}-2.55620 r^{15}+ r^{16}\nonumber\\
&& + 0.00329688 (2 - r)^2 r^{14} ln[1 - \frac{2}{r}]\},\\
\label{fr}
f(r)=
&&\frac{1}{r^{22}}\{3.76020-65.7119 r+518.234 r^2-2432.92 r^3+7549.78 r^4-16254.7 r^5\nonumber\\
&&+24782.9 r^6-26801.9 r^7+20235.5 r^8-10249.4 r^9+3179.84 r^{10}-457.973 r^{11}\nonumber\\
&&+7.15583 r^{12}-60.6217 r^{13}+81.1245 r^{14}-42.4149 r^{15}+7.02639 r^{16}+1.80243 r^{17}\nonumber\\
&&0.634747 r^{18}+0.0157998 r^{19}+1.11181 r^{20}-2.55590 r^{21}+ r^{22}\},\\
\label{Atr}
A_t(r)=
&&\frac{1}{r^{11}} \{0.00877198-0.0728020 r+0.226609 r^2-0.308562 r^3+0.144485 r^4\nonumber\\
&&+0.0102166 r^5-0.00262212 r^6+0.0244681 r^7+0.0581679 r^8+0.155568 r^9\nonumber\\
&&1.55343 r^{10}+1.88217*10^{-16} r^{11}+0.986741 r^{11} ln[1 -\frac{0.560861}{r}]\}.
\end{eqnarray}

\begin{figure}[H]
\centering
\subfigure[$a^*=0.280431$, $h_1^*=0.457715$]{
\label{comparisonfig_a} 
\includegraphics[scale=0.5]{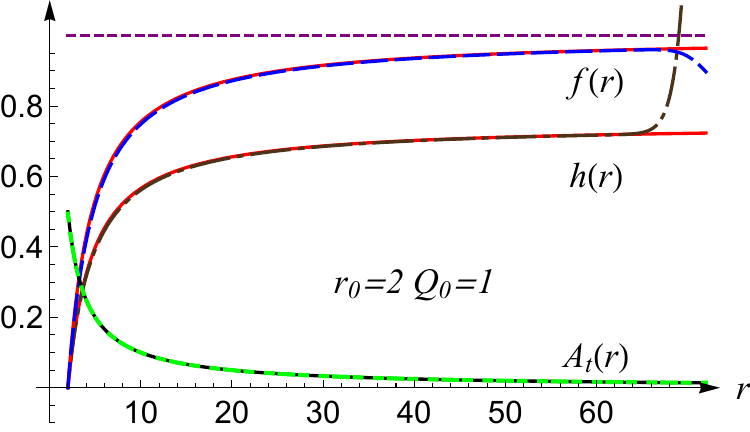}}
\quad
\subfigure[$a^*=0.174644$, $h_1^*=-0.212442$]{
\label{comparisonfig_b} 
\includegraphics[scale=0.5]{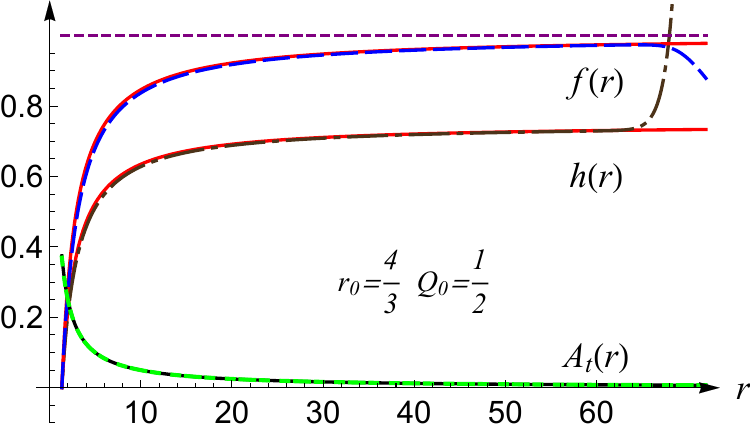}}
\caption{Comparison between analytical approximate solutions and numerical solutions for $h (r)$, $f (r)$, and $A_t (r)$. The solid line represents the analytical approximate solutions, and the dashed line represents the numerical solutions.}
\label{comparisonfig}
\end{figure}

The comparison between the analytical approximation solutions obtained by HAM and numerical solutions is plotted in Fig.~\ref{comparisonfig}.

\begin{figure}[H]
\centering
\label{accuracyfig_a} 
\includegraphics[scale=0.5]{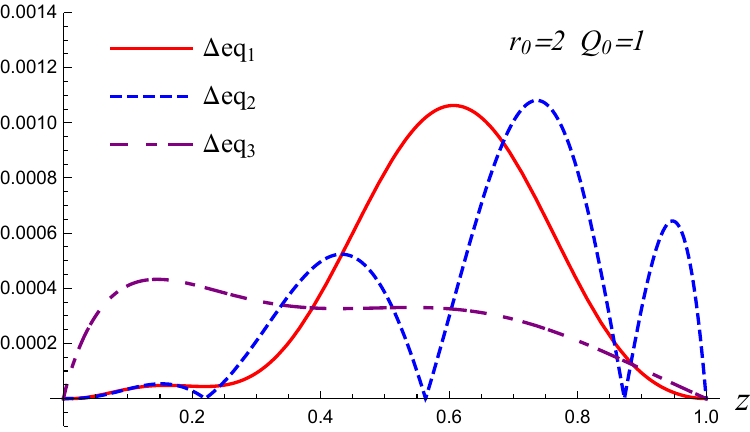}
\label{accuracyfig_b} 
\includegraphics[scale=0.5]{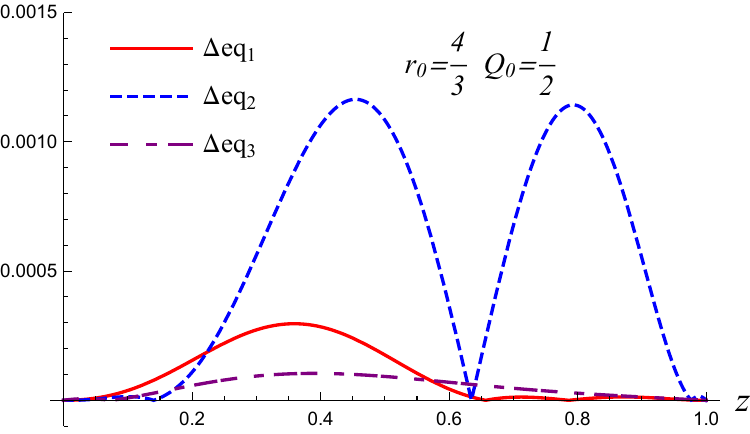}
\caption{Absolute errors for analytic approximate solutions from two field equations with different values of $r_0$ and $Q_0$ . Represented as the deviations of the analytical approximate solutions from the exact solutions. The parameters are the same as used in Fig.\ref{comparisonfig}.}
\label{accuracyfig}
\end{figure}

It is interesting to check the accuracy of the analytical approximate solutions. We plot these curves describing the deviations (Eqs.\eqref{delta eq1} and \eqref{delta eq2}) of the analytical approximate solutions from the exact solutions, as shown in Figure \ref{accuracyfig}. 

\begin{figure}[H]
\centering
\label{errorsfig_a} 
\includegraphics[scale=0.5]{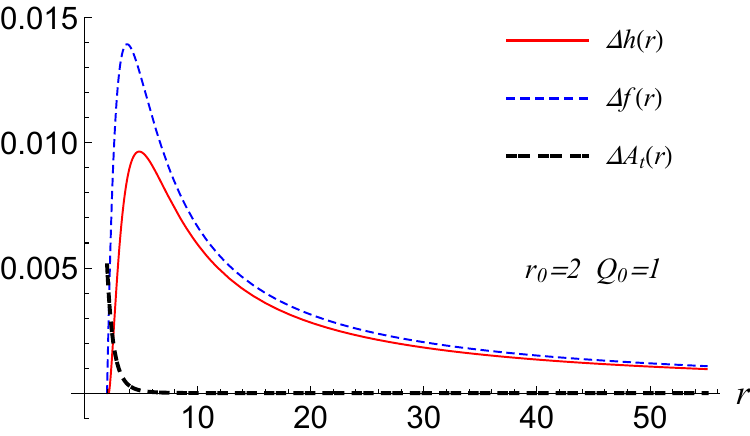}
\quad
\label{errorsfig_b} 
\includegraphics[scale=0.5]{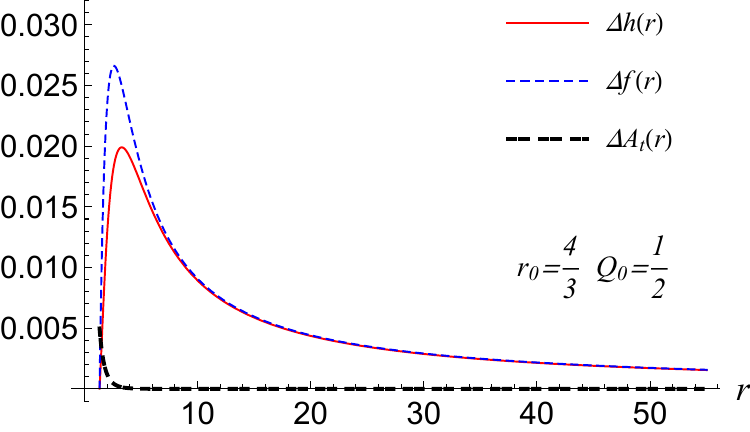}
\caption{The difference $\Delta h(r)$ (in red solid) and $\Delta f(r)$ (in blue dashed) between the analytic approximation solutions and numerical solutions. The parameters are the same as used in Fig.\ref{comparisonfig}.}\label{fig55}
\label{errorsfig}
\end{figure}

\begin{figure}[H]
\centering
\label{ralaerrorsfig_a} 
\includegraphics[scale=0.5]{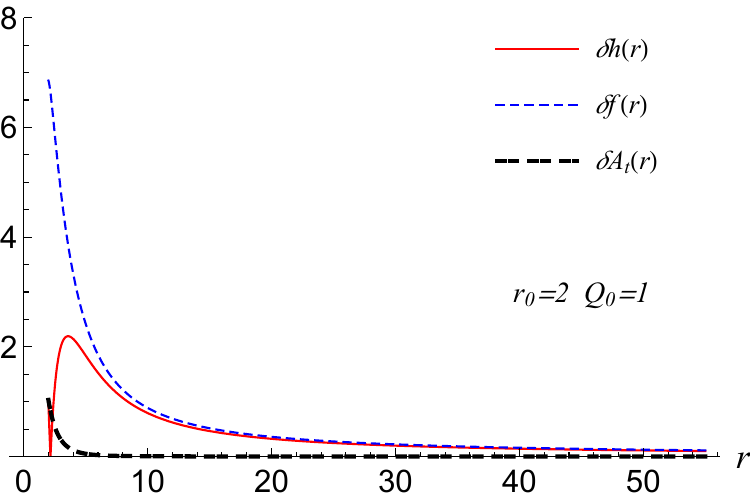}
\quad
\label{relaerrorsfig_b} 
\includegraphics[scale=0.5]{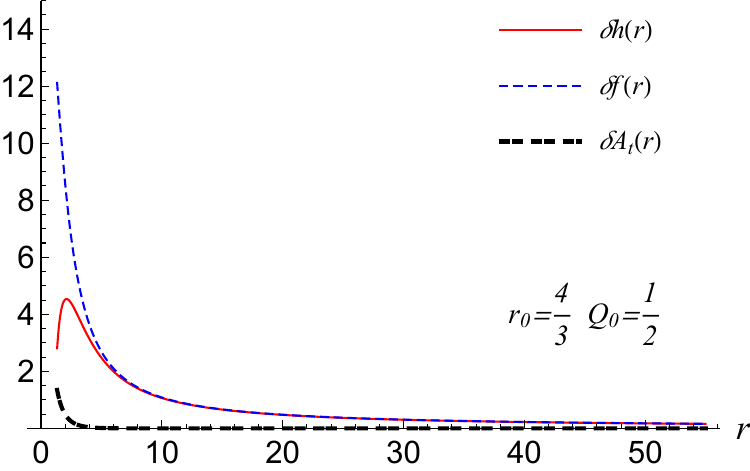}
\caption{The relative difference $\delta h(r)$ (in red solid) and $\delta f(r)$ (in blue dahsed) between the analytic approximation and numerical solutions. The parameters are the same as used in Fig.\ref{comparisonfig}.}\label{fig66}
\label{ralaerrorsfig}
\end{figure}

In order to compare with the numerical solutions, we will calculate the absolute errors between the analytical approximate and numerical solutions in Fig.~\ref{errorsfig}.
\begin{eqnarray}
\label{absolute errors}
&&\Delta h(r)=|h^{num}(r)-h^{ana}(r)]|, \quad \Delta f(r)=|f^{num}(r)-f^{ana}(r)]|,\nonumber\\
&&\Delta A_t(r)=|A_t^{num}(r)-A_t^{ana}(r)]|.
\end{eqnarray}
and the relative errors in Fig.~\ref{ralaerrorsfig}
\begin{eqnarray}
\label{relative errors}
&&\delta h(r)=\frac{|h^{num}(r)-h^{ana}(r)]|}{h^{num}(r)}\times100\%, \quad \delta f(r)=\frac{|f^{num}(r)-f^{ana}(r)]|}{f^{num}(r)}\times100\%,\nonumber\\
&&A_t(r)=\frac{|A_t^{num}(r)-A_t^{ana}(r)]|}{A_t^{num}(r)}\times100\%.
\end{eqnarray}
We find that main differences between numerical solutions and analytic approximation solutions occur close to the region
far from the black hole.

\section{Conclusions and discussions} \label{sec4}

In this paper, we discuss the Einstein-Weyl gravity coupled with Maxwell field. Even though the higher derivative curvature terms  are introduced into the gravitational action, the field equations always hold no more than second order derivative for the static and spherical metric functions. Then, we construct numerical solutions for charged black holes with fixed values of charge $Q_0$ and horizon radius $r_0$. Later, we turn to derive the analytical approximation solutions of charged black holes by using HAM in the Einstein-Maxwell-Weyl gravity. Instead of the convergence-control parameter $h$ obtained from so called ``h-curves'', here we calculate the optimal values of convergence parameter $h$ by evaluating the square residual error. Moreover, 
we further compare the analytic approximate solutions with the numerical solutions, and check these absolute and relative errors for these solutions, respectively. 

In addition, since the approximation is significantly accurate in the entire space-time outside the event horizon, it can be used for studying the properties of this particular black hole and the various phenomena. The present work is considered as an important work because we confirm that numerical solutions are consistent with an analytical approximate solution for the charged black holes.

 \vspace{1cm}

{\bf Acknowledgments}
 \vspace{1cm}

D. C. Z acknowledges financial support from the Initial Research Foundation of Jiangxi Normal  University. M. Z. acknowledges financial support from Natural Science Basic Research Program of Shaanxi (Program No.2023-JC-QN-0053). M. Y. L is supported by the Jiangxi Provincial Natural Science Foundation (Grant No. 20224BAB211020) .

\end{document}